\documentclass[aps,prx,amsmath,amssyb,floatfix,
twocolumn,superscriptaddress,10pt]{revtex4-1}
\usepackage{graphicx}
\usepackage{color}
\usepackage{calc}
\usepackage{hyperref}
\usepackage{braket}
\usepackage{bm}

\def\124{YBa$_2$Cu$_4$O$_8$ }

\def\C60{A$_x$C$_{60}$ }

\def\ie{{\it i.e.}}

\def\prl{{Phys. Rev. Lett. }}
\def\prb{{Phys. Rev. B }}

\def\be{\begin{equation}}
\def\ee{\end{equation}}
\def\ba{\begin{eqnarray}}
\def\ea{\end{eqnarray}}

\begin{document}

\title{Optimal inhomogeneity for pairing in Hubbard systems with
next-nearest-neighbor hopping}

\author{Gideon~Wachtel}
\affiliation{Racah Institute of Physics, The Hebrew University, Jerusalem 91904, Israel}
\affiliation{Department of Physics, University of Toronto, Toronto, Ontario M5S 1A7, Canada}
\author{Shirit~Baruch}
\affiliation{Racah Institute of Physics, The Hebrew University,
Jerusalem 91904, Israel}
\affiliation{Tel-Hai College, Upper Galilee 10120, Israel}
\author{Dror~Orgad}
\affiliation{Racah Institute of Physics, The Hebrew University,
Jerusalem 91904, Israel}

\date{\today}

\begin{abstract}

Previous studies have shown that bipartite Hubbard systems with
inhomogeneous hopping amplitudes can exhibit higher pair-binding
energies than the uniform model. Here we examine whether this result holds
for systems with a more generic band structure. To this end, we use exact
diagonalization and the density matrix renormalization group method to study
the $4\times 4$ Hubbard cluster and the two-leg Hubbard ladder with
checkerboard-modulated nearest-neighbor hopping, $t$, and next-nearest-neighbor
(diagonal) hopping, $t_d$. We find that the strongest pairing continues to occur at
an intermediate level of inhomogeneity. While the maximal pair-binding energy
is enhanced by a positive $t_d/t$, it is suppressed and appears at weaker repulsion
strengths and smaller hole concentrations when $t_d/t$ is negative.
We point out a possible connection between the pairing maximum and the magnetic
properties of the system.

\end{abstract}

\maketitle

\section{Introduction}

Consider two identical $N$-site clusters, each doped away from half filling by $M$
holes and residing in its ground state of energy $E_0(M)$. If holes tend to pair and
if $M$ is odd it is energetically preferable to move one hole between the clusters in
order to achieve a paired state on both. In this sense a positive pair-binding energy
\be
\label{eq:pbdef}
\Delta_{pb}(M/N)=2E_0(M)-E_0(M+1)-E_0(M-1),
\ee
serves as an indicator for pairing. Such evidence may be further strengthen by looking
at clusters whose ground state is a spin singlet. If the spin gap
\be
\label{eq:sgdef}
\Delta_s(M/N)=E_1(M)-E_0(M),
\ee
to the lowest $S=1$ excitation follows $\Delta_{pb}$ it is an indication that the
excitation is a result of a dissociation of a singlet hole pair into two separate holes.
In one dimension the relation is even more explicit as the opening of a spin gap entails
a non-zero amplitude for the superconducting order parameter \cite{our-review}.

Exact diagonalization studies have shown that a number of small Hubbard clusters
exhibit pair binding, which reaches a maximum at an intermediate strength of the
on-site repulsion \cite{Hubbard-clusters,Fye-clusters}. Similar behavior was observed
using the density-matrix renormalization group (DMRG) method in the two-leg Hubbard ladder,
where the binding energy is of the order of the spin gap and where both diminish with
doping \cite{Noack-ladder,Jeckelmann-ladder}.
These findings have inspired searches for superconductivity in two-dimensional systems composed
of coupled lower-dimensional building blocks, in hope of harnessing the pairing tendencies
of the latter. Such a strategy naturally gives rise to the question of what is the optimal level
of inhomogeneity for superconductivity \cite{Steve-optimal,Steve-handbook}.

Much of the research into the relationship between inhomogeneity and superconductivity from
repulsive interactions has been carried out using the plaquette Hubbard model \cite{Tsai1,Yao07}.
The model is constructed from $2\times 2$ plaquetes with on-site repulsion $U$ and nearest-neighbor
hopping $t$, where neighboring sites on different plaquettes are coupled by hopping $t'$.
Exact diagonalization of the $4\times 4$ site system \cite{Steve-exact} has found a substantial
maximum of the pair-binding energy at $t'/t\approx 0.5$, $U/t\approx 8$ and low hole doping.
A similar pairing maximum, occurring at intermediate inhomogeneity levels $t'/t\approx 0.5-0.7$
and interaction strengths $U/t\approx 5-8$, was subsequently found in larger plaquette systems
using the contractor renormalization (CORE) \cite{CORE} and DMRG \cite{Steve-DMRG} methods.
Furthermore, by calculating the other necessary ingredient for superconductivity, namely
phase stiffness, these studies have provided evidence that optimal inhomogeneity likely exists
also for the superconducting transition temperature, $T_c$, and not just for the pairing scale.
These findings were contested, however, by calculations using the dynamical cluster
approximation (DCA) \cite{Doluweera} and cellular dynamical mean-field theory (CDMFT)
\cite{Tremblay}, which have obtained a monotonic increase with $t'/t$ of both the $d$-wave
pairing interaction and of $T_c$ toward a maximum that is exhibited by the homogeneous model.
Nevertheless, a recent quantum Monte-Carlo (QMC) study \cite{Scalettar-plaquette} provides
support in favor of the CORE and DMRG findings. While the sign problem prevents
reliable calculation of $T_c$, it is manageable to low enough temperatures in order to
show that for $U/t=4$ the pairing vertex is most attractive at $t'/t\approx 0.4$ .
Finally, despite differences in details DCA \cite{Maier2010}, CDMFT \cite{Okamoto2010}
and QMC \cite{Scalettar-stripes} have all detected enhanced superconductivity in
Hubbard models with inhomogeneous charge density due to external potentials.

To date, optimal inhomogeneity for pairing, or more generally for superconductivity, has
been demonstrated only on the bipartite square lattice. It is therefore interesting to
explore the robustness of the phenomenon to changes in the band structure, not the least
because they are present in the cuprate high-temperature superconductors. Specifically,
the cuprates are often modeled using a tight-binding band structure of a square lattice
with hopping amplitudes that extend beyond the nearest-neighbor amplitude $t$. In particular,
it is necessary to include next-nearest-neighbor (diagonal) hopping $t_d$, with $t_d/t<0$,
to account for the observed Fermi surfaces \cite{Norman95}, and there are indications
that it plays a role in the physics governing $T_c$ of the hole-doped systems \cite{Pavarini01}.
The effects of such a term have been investigated both in the context of the Hubbard model
\cite{Riera,Tremblay95,Yamaji98,MaierDMFT00,GubernatisQMC01,Chen2012,Zheng2016} and its
strong-coupling descendent, the $t-J$ model
\cite{WhiteScalapino99,Shih04,Tohyama04,WhiteScalapino09,ScalapinoWhite12}.
It appears that different studies agree that various measures of pairing and superconductivity
are suppressed in the presence of $t_d/t<0$, at least for hole doping in the range $x<0.12$.
The results vary, however, for higher doping levels where calculations using DCA \cite{MaierDMFT00}
and DMRG \cite{WhiteScalapino99} continue to find suppression of superconductivity while density
matrix embedding theory \cite{Zheng2016} and variational QMC \cite{Shih04} indicate enhancement of
pairing correlations. A similar dichotomy also exists for positive $t_d/t$ where the first
group of methods finds enhanced superconductivity while the second yields an opposite trend.

Here we study the existence of optimal inhomogeneity for pairing in the plaquette Hubbard
model with diagonal hopping. To this end, we calculate the pair-binding energy and the spin gap
using exact diagonalization and DMRG. We show that pairing continues to peak at
intermediate levels of inhomogeneity but its strength depends on the sign of $t_d/t$.
Our results indicate that pairing is enhanced by the presence of $t_d/t>0$. On the
other hand, when $t_d/t<0$ pairing is suppressed for the higher hole concentrations examined
near $x=0.12$. It regains strength, however, at lower doping levels, or when $U/t$ is reduced.
We note that these effects can not be understood on the level of a single plaquette and speculate
on their possible connection to the magnetic properties of the system.

\section{Model and Results}

We consider the plaquette Hubbard model
\begin{equation}
\label{eq:H}
H=-\sum_{\langle i,j\rangle,\sigma}t_{ij}c^\dagger_{i\sigma}c_{j\sigma}
-\sum_{\langle\langle i,j\rangle\rangle,\sigma}t_{d,ij}c^\dagger_{i\sigma}c_{j\sigma}
+U\sum_i n_{i\uparrow}n_{i\downarrow},
\end{equation}
\begin{figure}[t]
  \centering
  \includegraphics[width=\linewidth,clip=true]{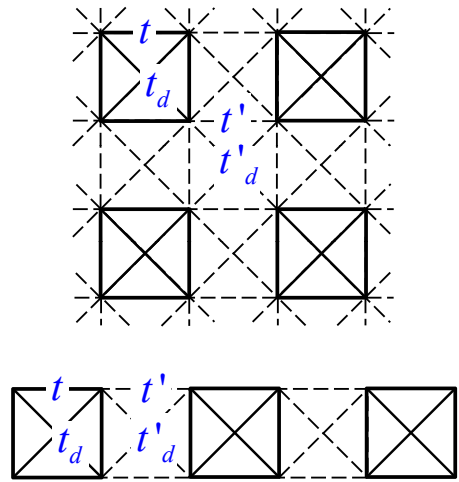}
  \caption{The $4\times 4$ cluster and a section of the two-leg ladder studied
  in this work.}
  \label{fig:diagladder}
\end{figure}
\begin{figure}[h!!!]
  \centering
  \includegraphics[width=\linewidth,clip=true]{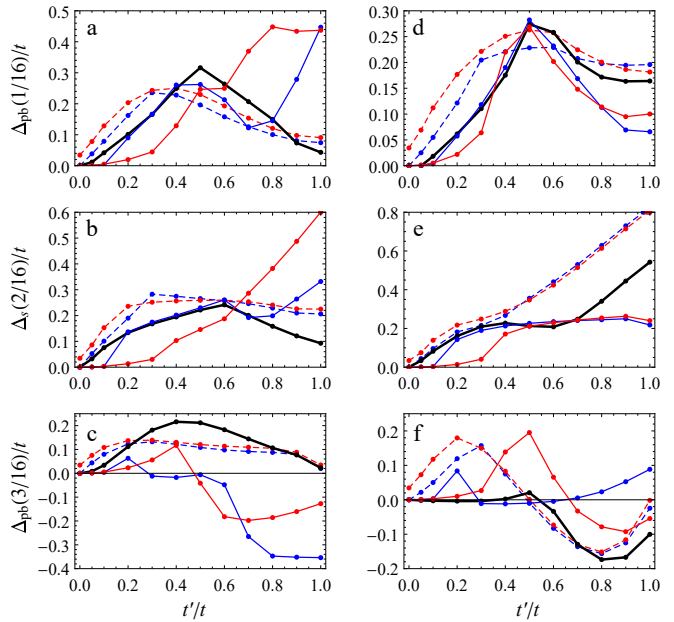}
  \caption{The $4\times 4$ cluster with $U/t=8$ and $t_d/t=-0.5$ (red), -0.3 (blue), 
  0 (black), 0.3 (dashed blue), 0.5 (dashed red).
  Left column: Results for the cluster with periodic boundary conditions. a) The pair-binding energy
  at 1/16 hole doping, b) The spin gap at 2/16 hole doping, c) The pair-binding energy at 3/16 hole
  doping. Right column (d-f): The corresponding quantities for the cluster with twisted boundary
  conditions. }
  \label{fig:cluster-comp}
\end{figure}

\noindent
where $c^\dagger_{i\sigma}$ creates an electron of spin polarization
$\sigma=\uparrow,\downarrow$ at site $i$, and $n_{i\sigma}=c^\dagger_{i\sigma}c_{i\sigma}$.
Here $\langle i,j\rangle$ and $\langle\langle i,j\rangle\rangle$ denote nearest-neighbor
and next-nearest-neighbor sites, respectively. The hopping amplitudes are modulated as
shown in Fig. \ref{fig:diagladder}. Neighboring sites within a plaquette are connected by
hopping $t$ while next-nearest-neighbors are connected by $t_d$. The corresponding amplitudes
across plaquette boundaries are $t'$ and $t_d'$. For simplicity we restrict ourselves to the
case in which the diagonal amplitudes are modulated with the same ratio as the nearest-neighbor
amplitudes, \ie, $t'_d/t_d=t'/t$. We have studied the model on the $4\times 4$ cluster and on
the two-leg ladder, depicted in Fig. \ref{fig:diagladder}. To obtain an estimate for the
finite-size effects in the smaller system we compare results for the cluster with periodic boundary
conditions in both directions: $(m+4,n)=(m,n+4)=(m,n)$ with results for a cluster subjected to twisted
boundary conditions: $(m+4,n)=(m,n)$ and $(m,n+4)=(m+2,n)$. We use open boundary conditions
for the two-leg ladder.

\begin{figure}[t!!!]
  \centering
  \includegraphics[width=\linewidth,clip=true]{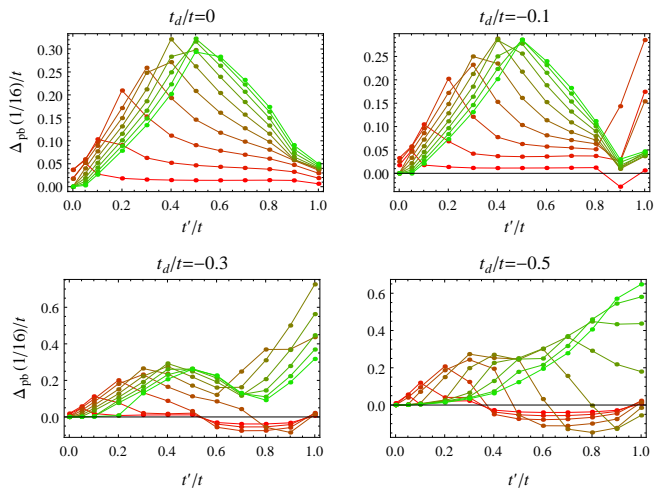}
  \caption{The pair-binding energy of the $4\times 4$ cluster with periodic boundary conditions
  at hole doping 1/16 and various $t_d/t$. The results are for $U/t=1 ({\rm red}) ,2,\cdots ,10 ({\rm green})$.}
  \label{fig:cluster-U}
\end{figure}

The pair-binding energy and the spin gap for the smallest available hole concentrations on
the $4\times 4$ cluster were calculated using exact diagonalization.
They are depicted in Fig. \ref{fig:cluster-comp} for the case $U/t=8$. We find that the
pair-binding energy of the $x=1/16$ system is largely insensitive to changes in the
boundary conditions over a range of $t'/t$ that shrinks with increasing $|t_d/t|$.
For $|t_d/t|\leq 0.1$ (not shown) the two sets of results follow each other and differ by at most 20\%
as long as $0\leq t'/t\lesssim 0.8$. In particular, a clear maximum in $\Delta_{pb}$ is
observed at $t'/t=0.5$. This maximum is also present in the $t_d=-0.3 t$ data, but due
to the increased sensitivity to the boundary conditions above $t'/t\approx 0.65$ we are
unable to determine whether it constitutes a global pairing maximum. The even larger
sensitivity of the $t_d=-0.5 t$ results precludes reaching a conclusion about the
existence of optimal inhomogeneity for pairing in this case. At the same time, the results 
for positive $t_d$ are more robust and $\Delta_{pb}(1/16)$ exhibits a consistent maximum 
around $t'/t=0.3-0.5$ with a clear enhancement compared to the $t_d=0$ case for inhomogeneity 
levels below the maximum. The evolution of
$\Delta_{pb}$ with $U/t$ for $t_d<0$ is presented in Fig. \ref{fig:cluster-U}. Evidently, within
the range of $t'/t$ discussed above the position and the magnitude of the maximal binding
energy increases with $U/t$ until it reaches a global maximum around $U/t\approx 8-10$.

Positive pair-binding energy may also be associated with a tendency of the system to phase
separate. In order to distinguish between pairing and phase separation one needs
to calculate the surface tension between the hole-rich and hole-poor phases \cite{Hellberg97}.
A cruder way is to look for negative inverse compressibility, as a sign for instability
towards phase separation. We, however, always find its discrete version
\begin{equation}
\label{eq:compress}
\kappa^{-1}\propto E_0(M+2)+E_0(M-2)-2E_0(M),
\end{equation}
\begin{figure}[t!!!]
  \includegraphics[width=\linewidth]{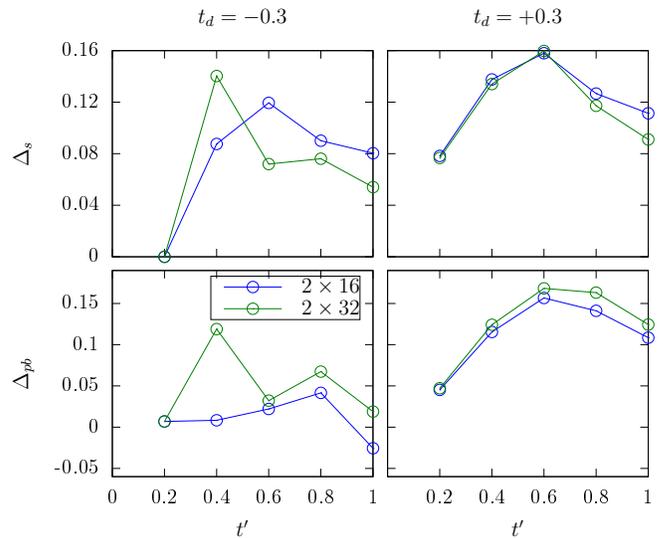}
  \caption{The spin gap and pair-binding energy of the 2-leg plaquette
   ladder with $U=8t$ and hole doping $x=1/16$.}
  \label{fig:U8n/16}
\end{figure}
\begin{figure}[h!!!]
  \includegraphics[width=\linewidth]{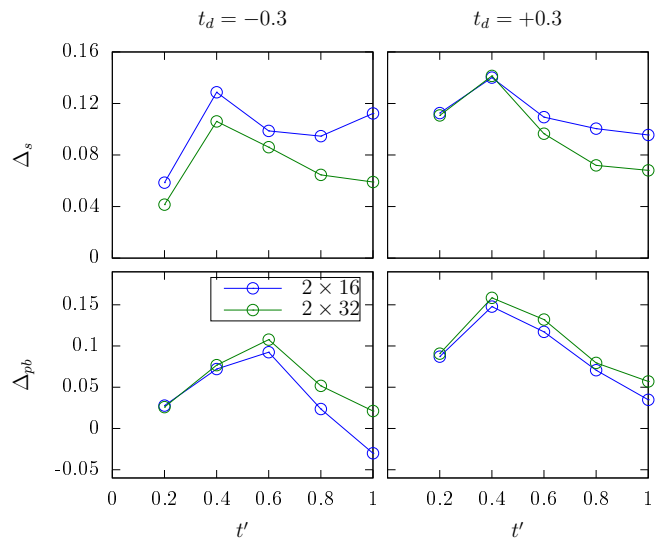}
  \caption{The spin gap and pair-binding energy of the 2-leg plaquette
   ladder with $U=4t$ and hole doping $x=1/16$.}
  \label{fig:U4n/16}
\end{figure}
to be positive. Further support for pairing comes from the fact that the spin gap of the
system with two doped holes roughly follows $\Delta_{pb}(1/16)$, as presented in Fig. \ref{fig:cluster-comp}.
In contrast, $\Delta_{pb}(3/16)$ exhibits large sensitivity to the boundary conditions and we can not
determine whether holes pair on the cluster at this higher doping level.

\begin{figure}[t!!!]
  \includegraphics[width=\linewidth]{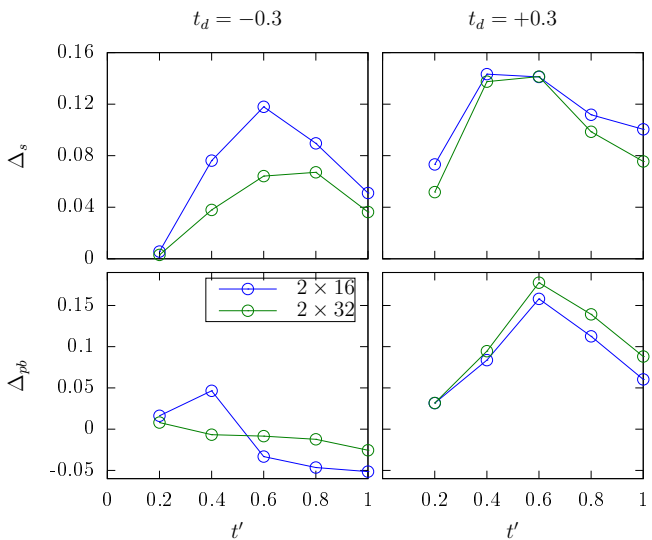}
  \caption{The spin gap and pair-binding energy of the 2-leg plaquette
   ladder with $U=8t$ and hole doping $x=1/8$.}
  \label{fig:U8n1/8}
\end{figure}

\begin{figure}[h!!!]
  \includegraphics[width=\linewidth]{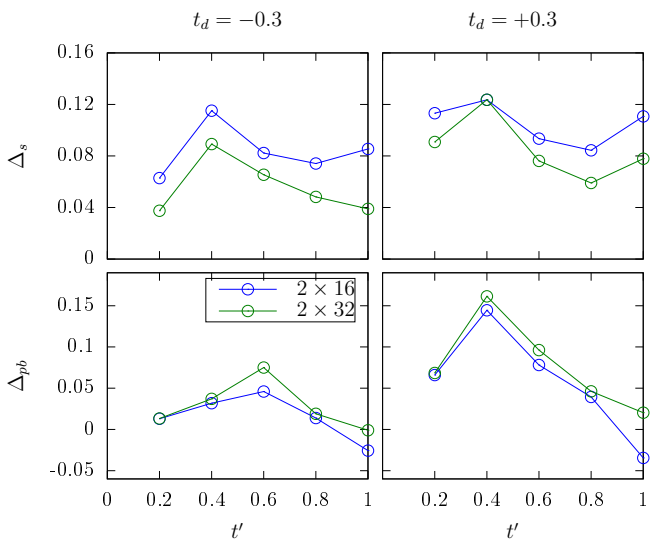}
  \caption{The spin gap and pair-binding energy of the 2-leg plaquette
   ladder with $U=4t$ and hole doping $x=1/8$.}
  \label{fig:U4n1/8}
\end{figure}

In an effort to substantiate the exact diagonalization study we have used DMRG
to calculate $\Delta_s$ and $\Delta_{pb}$ of the two-leg plaquette ladder
with $t_d/t=\pm0.3$. During the calculation we have truncated the density
matrix, keeping up to about 3200 states in order to reach low enough truncation
errors. The relatively large number of kept states (larger than needed when $t_d=0$ \cite{Steve-DMRG}) 
meant we could deal with ladders of up to $2\times32$ sites. Our results for the ladder with hole 
doping $x=1/16$ are summarized in Figures \ref{fig:U8n/16} and \ref{fig:U4n/16}, and for $x=1/8$
in Figures \ref{fig:U8n1/8} and \ref{fig:U4n1/8}. All quantities are given in units of $t$.


Concentrating first on the $t_d/t=-0.3$ case, we find general agreement between
the DMRG results and those obtained using exact diagonalization. Specifically,
a positive pair-binding energy, accompanied by $\kappa^{-1}>0$, is observed for $x=1/16$
and exhibits a peak at an intermediate inhomogeneity level. The peak is robust in the
$U/t=4$ ladder, where the results change little upon increasing the length of the system.
On the other hand, the results for $U/t=8$ still show substantial size dependence in going
from the $2\times 16$ to the $2\times 32$ ladder. Nevertheless, the tendency of $\Delta_{pb}$
to grow with the system size and the fact that it shows similar features to $\Delta_s$
make it plausible that optimal inhomogeneity for pairing also exists in the thermodynamic limit.
Increasing the hole concentration to $x=1/8$ leads to a $\Delta_{pb}$ that is indiscernible
when $U/t=8$. This stands in contrast to the $t_d=0$ ladder under similar conditions
where $\Delta_{pb}$ attains a maximal value of about $0.15t$ \cite{Steve-DMRG}.
Optimal pairing reappears upon lowering the interaction strength to $U/t=4$
but it is still somewhat weaker than its value when $t_d=0$ \cite{Steve-DMRG}. We therefore
conclude that while optimal inhomogeneity continues to exist in the presence of $t_d/t<0$,
such a hopping term tends to reduce the optimal pairing scale, particularly for stronger
interactions and higher hole concentrations. On the contrary, our results clearly show that
a next-nearest-hopping term with $t_d/t>0$ enhances the pairing maximum, for all values of
$U/t$ and $x$ studied by us.

\section{Discussion}

\begin{figure}[t!!!]
  \centering
  \includegraphics[width=\linewidth,clip=true]{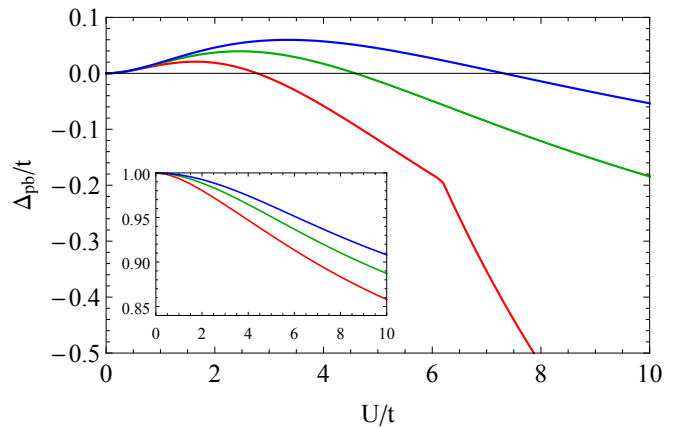}
  \caption{The pair-binding energy of the $2\times 2$ plaquette as function of  $U/t$ for
  $t_d/t=-0.3$ (red), 0 (green), and 0.3 (blue). The inset depicts the $U/t$ dependence of
  the overlap between the two-hole ground state and the normalized state obtained by applying
  the pair annihilation operator $P_{12}-P_{23}+P_{34}-P_{41}$ to the undoped ground state.}
  \label{fig:square-pbe}
\end{figure}

What might be the origin of the pairing maximum and its dependence on $t_d$? One may try to look for the
reasons in the Hubbard plaquette itself. Fig. \ref{fig:square-pbe} shows, however, that the pair-binding
energy of the square is already negative at the $U/t$ values which correspond to the pairing maximum
observed in the extended systems. Therefore, the latter is not a single plaquette effect. Nevertheless,
$\Delta_{pb}$ of the large and small systems share some characteristics, such as its tendency to decrease
as one moves from positive to negative $t_d/t$. It has been suggested in the context of the $t-J$ model
\cite{WhiteScalapino09}, that this may be due to the fact that negative $t_d/t$ is less favorable for
creation of paired states with $d$-wave symmetry.

It can be shown that for $|t_d/t|<1$ the ground state of the undoped plaquette, $|N_h=0\rangle$,
is an $S=0$ singlet which is odd under $\pi/2$ rotations, and approaches the "RVB" state
$(1/\sqrt{12})(P^\dagger_{12}P^\dagger_{34}-P^\dagger_{14}P^\dagger_{23})|0\rangle$ at large $U/t$.
Here, $P^\dagger_{ij}=c^\dagger_{i\uparrow}c^\dagger_{j\downarrow}+c^\dagger_{j\uparrow}c^\dagger_{i\downarrow}$
creates a singlet electron pair on sites $i$ and $j$ (numbered sequentially around the square).
When the interactions are weak the two holes ground state, $|N_h=2\rangle$, is generated from $|N_h=0\rangle$
by $P_{12}-P_{23}+P_{34}-P_{41}$. However, as $U/t$ is made stronger the pairing operator includes
terms that create holes on next-nearest-neighbor sites as well \cite{Dagotto92,Poilblanc94}.
We find that the effect of these terms increases when $t_d/t$ turns from positive to negative,
see the inset  of Fig. \ref{fig:square-pbe}, but for all values $|t_d/t|<1$ the state $|N_h=2\rangle$
remains a spin singlet that lies in the identity representation of the rotation group.
Therefore, the pair annihilation operator connecting $|N_h=0\rangle$ and $|N_h=2\rangle$ must transform as
$d_{x^2-y^2}$. In this sense the inclusion of diagonal hopping does not affect the pairing symmetry
on the square. Nevertheless, it does affect the energetics of the pairing process.
Interestingly, both $|N_h=0\rangle$ and its energy are independent of $t_d$ (in the range considered here).
At the same time, the energies of $|N_h=2\rangle$ and of the single hole ground state $|N_h=1\rangle$
increase with $U/t$ in a rate which depends on $t_d$. We find that the suppression of the plaquette $\Delta_{pb}$
at large $U/t$ and $t_d/t<0$ is largely driven by the slower increase of the single hole energy.
In particular, for $-1<t_d/t<0.25$ there exists a critical $U/t$ (that increases with $t_d$)
where $|N_h=1\rangle$ turns from a degenerate quartet with $S_z=\pm 1/2$ and plaquette
momentum $(0,\pi)$ or $(\pi,0)$, to a degenerate quartet with constant energy $-2t+t_d$,
made of $S_z=\pm1/2$ and $S_z=\pm3/2$ doublets with momentum $(\pi,\pi)$, thereby leading to the break
seen in Fig. \ref{fig:square-pbe}.

As noted above, while some features of the $t_d$ dependence of $\Delta_{pb}$ follow the behavior found
on the single plaquette level, the existence of a pairing maximum as function of $t'/t$ can not be
understood from such considerations. Instead, we would like to point out a correlation between the
pairing maximum and the magnetic properties of the system. It is well known that in the large $U/t$ limit
the half filled Hubbard model maps onto the $S=1/2$ Heisenberg model with $J=4t^2/U$ \cite{Assa}. Correspondingly,
the plaquette Hubbard model maps onto the plaquette Heisenberg model with $J'/J=(t'/t)^2$. Since the
ground state of the uniform model ($J'=J$) exhibits N${\rm \acute{e}}$el antiferromagnetic (AFM) long-range order,
and that of the disconnected system ($J'=0$) is a product of RVB states on individual plaquettes, one expects
that a quantum critical point (QCP) separates the two at an intermediate $J'/J$. Such expectation has
been borne out by numerical calculations \cite{Capponi04,Wenzel09} which find a QCP at $J'/J\simeq 0.55$.
A recent QMC study of the plaquette Hubbard model \cite{Scalettar-plaquette} provides evidence that this
QCP survives at half filling for lower values of the interaction strength. Interestingly, the observed
QCP at $t'/t\approx 0.5,0.6$ for $U/t=4,8$ resides in the vicinity of the $t'/t$ value for which the
product of the pairing vertex and the uncorrelated pairing susceptibility is closest to -1, where a
superconducting instability would develop. This coincidence joins a related behavior which we have
noticed in our CORE study \cite{CORE} of the model away from half-filling. While no transition to
long-range AFM order is observed (nor expected beyond small hole doping levels), the maximal $\Delta_{pb}$
does occur at $t'/t$ around which AFM correlations build up from the RVB background. It is therefore
possible that the enhanced magnetic fluctuations generated by the inhomogeneity-induced QCP,
or its related crossover at finite doping, are the mediator responsible for the enhanced pairing.

The presence of diagonal hopping turns the large-$U$ limit of the half-filled Hubbard model into the
$J_1-J_2$ Heisenberg model with AFM couplings satisfying $J_2/J_1=(t_d/t)^2$. The frustration introduced
by the next-nearest-neighbor AFM coupling, $J_2$, causes a sequence of quantum phase transitions where
N${\rm \acute{e}}$el order is first lost at $J_2/J_1\approx 0.4$, in favor of a nonmagnetic state that
is either columnar or plaquette valence-bond-solid, which then gives way to a collinear (striped) magnetic
order at $J_2/J_1\approx 0.6$ \cite{Dagotto89,Sushkov01,Fisher14}. Variational cluster approximations
of the half-filled Hubbard model with positive $t_d/t$ point at a similar picture where a nonmagnetic phase
in the range $0.7\lesssim t_d/t\lesssim 0.8$ separates the magnetically ordered states down to $U/t\approx 5$,
where it spreads out \cite{Tremblay08,Yamada13}. The fact that these transitions occur at relatively large
values of $t_d/t$ suggests that they do not play a role in establishing the results presented by us, or in
the cuprate superconductors. Nevertheless, the extent of the nonmagnetic phase grows in the plaquette
$J_1-J_2$ Heisenberg model \cite{Singh99,Gotze12}. Ref. \onlinecite{Gotze12} predicts that for $J_2/J_1=0.1$
a transition to a N${\rm \acute{e}}$el state takes place once the intra-plaquette couplings are about twice
the inter-plaquette ones. This would correspond to a transition in the large-$U$ limit of the
half-filled plaquette Hubbard model with $|t_d/t|=0.3$ at $t'/t\approx 0.7$, not too far from where we
observe the pairing maximum in the system with $t_d/t=0.3$ and $U/t=8$. However, this observation does
not explain why the pairing maximum seems to appear at lower values of $t'/t$ when $t_d/t=-0.3$. To
answer this question and strengthen the conjectured tie between pairing and a magnetic QCP further
study of the doped plaquette Hubbard model is called for.

\acknowledgments
This research was supported by the United States-Israel Binational
Science Foundation (Grant No. 2014265) and by the Israel Science Foundation (Grant No. 585/13).
GW additionally acknowledges support from NSERC of Canada and the CIfAR Quantum Materials Programme.


\begin{thebibliography}{999}

\bibitem{our-review} E.~W.~Carlson, V.~J.~Emery, S.~A.~Kivelson,
and D.~Orgad, in {\it Superconductivity: Novel Superconductors},
edited by K.~H.~Bennemann and J.~B.~Ketterson (Springer-Verlag, Berlin, 2008),
Vol 2, p. 1225.

\bibitem{Hubbard-clusters} S.~R.~White, S.~Chakravarty, M.~P.~Gelfand, and S.A.~Kivelson,
\prb {\bf 45}, 5062 (1992).

\bibitem{Fye-clusters} R.~M.~Fye, D.~J.~Scalapino, and R.~T.~Scalettar, \prb {\bf 46},
8667 (1992).

\bibitem{Noack-ladder} R.~M.~Noack, S.~R.~White, and D.~J.~Scalapino, Physica C
{\bf 270}, 281 (1996).

\bibitem{Jeckelmann-ladder} E.~Jeckelmann, D.~J.~Scalapino, and S.~R.~White,
\prb {\bf 58}, 9492 (1998).


\bibitem{Steve-optimal} E.~Arrigoni and S.~A.~Kivelson, \prb {\bf 68}, 180503(R) (2003).

\bibitem{Steve-handbook} S.~A.~Kivelson and E.~Fradkin, in
{\it Handbook of High-Temperature Superconductivity}, edited by
J.~R.~Schrieffer and J.~S.~Brooks (Springer, New York, 2007).

\bibitem{Tsai1} W.~F.~Tsai and S.~A.~Kivelson, \prb {\bf 73}, 214510 (2006).

\bibitem{Yao07} H.~Yao, W.-F.~Tsai, and S.~A.~Kivelson, \prb {\bf 76},
161104(R) (2007).

\bibitem{Steve-exact} W.-F.~Tsai, H.~Yao, A.~L${\rm \ddot{a}}$uchli, and S.~A.~Kivelson,
\prb {\bf 77}, 214502 (2008).

\bibitem{CORE} S.~Baruch and D.~Orgad, \prb {\bf 82}, 134537 (2010).

\bibitem{Steve-DMRG} G.~Karakonstantakis, E.~Berg, S.~R.~White, and
S.~A.~Kivelson, \prb {\bf 83}, 054508 (2011).

\bibitem{Doluweera} D.~G.~S.~P.~Doluweera, A.~Macridin, T.~A.~Maier, M.~Jarrell,
and T.~Pruschke, \prb {\bf 78}, 020504(R) (2008).

\bibitem{Tremblay} S.~Chakraborty, D.~S$\acute{\rm e}$n$\acute{\rm e}$chal,
and A.-M.~S.~Tremblay, \prb {\bf 84}, 054545 (2011).

\bibitem{Scalettar-plaquette} T.~Ying, R.~Mondaini, X.~D.~Sun, T.~Paiva,
R.~M.~Fye, and R.~T.~Scalettar, \prb {\bf 90}, 075121 (2014).

\bibitem{Maier2010} T.~A.~Maier, G.~Alvarez, M.~Summers, and T.~C.~Schulthess,
\prl {\bf 104}, 247001 (2010).

\bibitem{Okamoto2010} S.~Okamoto and T.~A.~Maier, \prb {\bf 81}, 214525 (2010).

\bibitem{Scalettar-stripes} R.~Mondaini, T.~Ying, T.~Paiva, and R.~T.~Scalettar,
\prb {\bf 86}, 184506 (2012).

\bibitem{Norman95} M.~R.~Norman, M.~Randeria, H.~Ding, and J.~C.~Campuzano,
\prb {\bf 52}, 615 (1995).

\bibitem{Pavarini01} E.~Pavarini, I.~Dasgupta, T.~Saha-Dasgupta, O.~Jepsen,
and O.~K.~Andersen, \prl {\bf 87}, 047003 (2001).





\bibitem{Riera} J.~A.~Riera, \prb {\bf 40}, 833 (1989).


\bibitem{Tremblay95} A.~F.~Veilleux, A.-M.~Dar${\rm \acute{e}}$, L.~Chen,
Y.~M.~Vilk, and A.-M.~S.~Tremblay, \prb {\bf 52}, 16255 (1995).

\bibitem{Yamaji98} K.~Yamaji, T.~Yanagisawa, T.~Nakanishi, and S.~Koike,
Physica C {\bf 304}, 225 (1998).


\bibitem{MaierDMFT00} T.~Maier, M.~Jarrell, T.~Pruschke, and
J.~Keller, \prl {\bf 85}, 1524 (2000).

\bibitem{GubernatisQMC01} Z.~B.~Huang, H.~Q.~Lin, and
J.~E.~Gubernatis, \prb {\bf 64}, 205101 (2001).


\bibitem{Chen2012} K.-S.~Chen, Z.~Y.~Meng, T.~Pruschke, J.~Moreno, and
M.~Jarrell, \prb {\bf 86}, 165136 (2012).

\bibitem{Zheng2016} B.-X.~Zheng and G.~K.-L.~Chan, \prb {\bf 93},
035126 (2016).




\bibitem{WhiteScalapino99} S.~R.~White and D.~J.~Scalapino,
\prb {\bf 60}, R753 (1999).

\bibitem{Shih04} C.~T.~Shih, T.~K.~Lee, R.~Eder, C.-Y.~Mou, and
Y.~C.~Chen, \prl {\bf 92}, 227002 (2004).

\bibitem{Tohyama04} T.~Tohyama, \prb {\bf 70}, 174517 (2004).

\bibitem{WhiteScalapino09} S.~R.~White and D.~J.~Scalapino,
\prb {\bf 79}, 220504(R) (2009).

\bibitem{ScalapinoWhite12} D.~J.~Scalapino and S.~R.~White,
Physica C {\bf 481}, 146 (2012).

\bibitem{Hellberg97} C.~S.~Hellberg and E.~Manousakis, \prl {\bf 78},
4609 (1997).

\bibitem{Dagotto92} E.~Dagotto, J.~Riera, and D.~Scalapino, \prb {\bf 45},
5744 (1992).

\bibitem{Poilblanc94} D.~Poilblanc, \prb {\bf 49}, 1477 (1993).

\bibitem{Assa} A.~Auerbach, {\it Interacting Electrons and Quantum Magnetism}
(Springer, New York, 1994).

\bibitem{Capponi04} S.~Capponi, A.~L${\rm \ddot{a}}$uchli, and M.~Mambrini,
\prb {\bf 70}, 104424 (2004).

\bibitem{Wenzel09} S.~Wenzel and W.~Janke, \prb {\bf 79}, 014410 (2009).

\bibitem{Dagotto89} E.~Dagotto and A.~Moreo, \prb {\bf 39}, 4744 (1989).

\bibitem{Sushkov01} O.~P.~Sushkov, J.~Oitmaa, and Z.~Weihong, \prb {\bf 63}, 104420 (2001).

\bibitem{Fisher14} S.-S.~Gong, W.~Zhu, D.~N.~Sheng, O.~I.~Motrunich,
and M.~P.~A.~Fisher, \prl {\bf 113}, 027201 (2014).


\bibitem{Tremblay08} A.~H.~Nevidomskyy, C.~Scheiber, D.~S$\acute{\rm e}$n$\acute{\rm e}$chal,
and A.-M.~S.~Tremblay, \prb {\bf 77}, 064427 (2008).

\bibitem{Yamada13} A.~Yamada, K.~Seki, R.~Eder, and Y.~Ohta, \prb {\bf 88}, 075114 (2013).

\bibitem{Singh99} R.~R.~P.~Singh, Z.~Weihong, C.~J.~Hamer, and J.~Oitmaa, \prb {\bf 60},
7278 (1999).

\bibitem{Gotze12} O.~G${\rm \ddot{o}}$tze, S.~E.~Kr${\rm \ddot{u}}$ger, F.~Fleck,
J.~Schulenburg, and J.~Richter, \prb {\bf 85}, 224424 (2012).



%
%
%
%
%
%
%
%
%
%
%
%
%
%



\end{thebibliography}
\end{document}